# Parallel Matrix Condensation for Calculating Log-Determinant of Large Matrix

Xiaomeng Dong[1], EN Barnett[1], Sudarshan K.Dhall[1]

## Abstract

Calculating the log-determinant of a matrix is useful for statistical computations used in machine learning, such as generative learning which uses the log-determinant of the covariance matrix to calculate the log-likelihood of model mixtures. The log-determinant calculation becomes challenging as the number of variables becomes large. Therefore, finding a practical speedup for this computation can be useful. In this study, we present a parallel matrix condensation algorithm for calculating the log-determinant of a large matrix. We demonstrate that in a distributed environment, Parallel Matrix Condensation has several advantages over the well-known Parallel Gaussian Elimination. The advantages include high data distribution efficiency and less data communication operations. We test our Parallel Matrix Condensation against self- implemented Parallel Gaussian Elimination as well as ScaLAPACK (Scalable Linear Algebra Package) on 1000 x1000 to 8000x8000 for 1,2,4,8,16,32,64 and 128 processors. The results show that Matrix Condensation yields the best speed-up among all other tested algorithms. The code is available on https://github.com/vbvg2008/MatrixCondensation

## 1. Introduction

Matrix condensation originates from Dodgson's condensation [2], where the determinant of an $N \times N$ matrix can be condensed into the determinant of an *N-1× N-1* matrix. One limitation that hampers the robustness of this method is that it requires non-zero elements in the interior of matrix

[1]University of Oklahoma, (Xiaomeng.dong-1, en.barnett-1, sdhall)@ou.edu

[1, 2, 3]. The non-zero interior limitation was later overcome by the work of Salem and Said [4], in which they modify the condensation method and generalize it into a different form:

$$\det(A) = \det(B) / a_{k,l}^{N-2} \quad for\ N > 2 \tag{1}$$

Where $A$ is an $N \times N$ matrix, $a_{k,l}$ is the element on row $k$ and column $l$ of $A$. $B$ is an $N\text{-}1 \times N\text{-}1$ matrix with elements being calculated as follows:

$$B_{i,j} = \begin{cases} Det\left(\begin{bmatrix} a_{i,j} & a_{i,l} \\ a_{k,j} & a_{k,l} \end{bmatrix}\right) & if\ i < k, j < l \\[6pt] Det\left(\begin{bmatrix} a_{k,l} & a_{k,j+1} \\ a_{i,l} & a_{i,j+1} \end{bmatrix}\right) & if\ i < k, j \geq l \\[6pt] Det\left(\begin{bmatrix} a_{k,l} & a_{k,j} \\ a_{i+1,l} & a_{i+1,j} \end{bmatrix}\right) & if\ i \geq k, j < l \\[6pt] Det\left(\begin{bmatrix} a_{k,l} & a_{k,j+1} \\ a_{i+1,l} & a_{i+1,j+1} \end{bmatrix}\right) & if\ i \geq k, j \geq l \end{cases} \tag{2}$$

Note that in Equations (1) and (2), both $k$ and $l$ can be arbitrarily chosen from 1 to $N$. Later studies [1, 3] have all assumed that k starts from 1, which means that the pivot will always be chosen from the first row during the condensation process. However, we are going to demonstrate in Section 2.1 that keeping k arbitrary will provide better convenience and flexibility for parallel implementation.

In order to avoid the division in Equation (1), the condensation method is further improved by Haque and Maza [3]. The division can be ignored when $a_{k,l}$ is 1, which can be achieved by factoring out $a_{kl}$ from column $l$. Then Equation (1) can be written as:

$$\det(A) = a_{kl} \det(A^*) = a_{kl} \det(B^*) \tag{3}$$

Where $A^*$ is a matrix that has $a_{k,l}$ factored out from column $l$ and $B^*$ is the matrix computed from $A^*$ using Equation (2). Since an entire column will be divided by the pivot $a_{k,l}$, one important step for this method is choosing the column index $l$ to avoid division by zero. Beliakov[1] defined $l$ to be the smallest column index of non-zero element on $k^{th}$ row of A. However, the accuracy of machine floating point numbers can be at risk when encountering an extremely small yet non-zero pivot. Haque[3] goes further by taking floating point digits into consideration and choosing $l$ to be the column index with a non-zero element closest to 1. In our method, different from previous studies, we propose using the column with the maximum absolute value. This slight modification can give us more robustness with the same computational cost as will be explained in detail in Section 2.2. Another noteworthy comment is that all studies mentioned above [1,3] factored $a_{k,l}$ out of column $l$. In fact, Equation (3) holds when factoring $a_{k,l}$ from row $k$ as well. In parallel implementation, this seemingly trivial property of matrix condensation can save a lot of data communication between processors when doing partial pivoting, as we will illustrate in section 2.3.

## 2. Parallel Matrix Condensation

### 2.1 Choosing an arbitrary pivoting row

Assuming the parallel algorithm is developed for row-major distributed memory architecture, we have matrix A with size 16x16 and four processors P1 to P4. The data of A can be distributed either by block data distribution or cyclic data distribution (Figure 1). If the pivot is always chosen from top to bottom during the condensation process as indicated by previous studies[1,3], then only cyclic data distribution will provide load-balancing whereas block data distribution is unbalanced (P1 will be idle after 4 condensations).

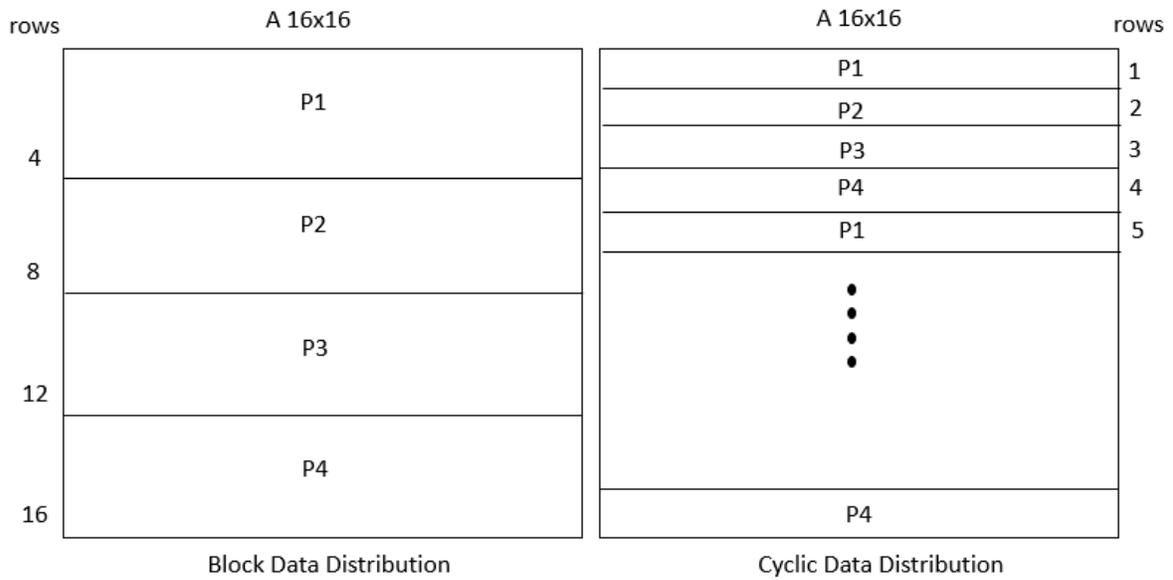

*Figure 1 Block Data Distribution V.S. Cyclic Data Distribution*

However, as mentioned earlier, Equation (2) holds true for an arbitrarily chosen pivot row *k*, which can free us from the top-to-bottom restriction for the condensation process. Therefore, by making use of this property, matrix condensation can achieve load-balancing while using block data distribution as shown in Figure 2.

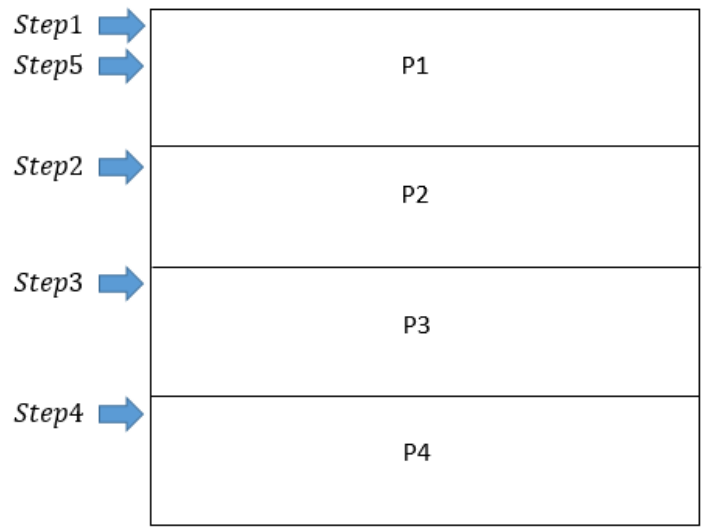

*Figure 2 Matrix Condensation Steps by Block Data Distribution*

Matrix condensation can gain better data distribution efficiency by making use of block data distribution scheme, which has certain advantages over cyclic data distribution: 1. the block data distribution is more easily achieved whereas cyclic data distribution requires special mapping. 2. The data chunks being distributed are contiguous in memory, therefore, it is faster and easier to distribute among nodes.

## 2.2 Choosing pivot column for partial pivoting

Once the pivot row is selected, the pivot column needs to be carefully chosen to ensure the pivot is non-zero. Moreover, as is pointed out in [3], we are expected to reduce the potential of overflow when dividing by the pivot. Differing from other methods [1, 3], our method chooses the pivot column to be the column that has the maximum absolute value. Compared with choosing the pivot that is closest to 1, our seemingly trivial modification provides much more robustness. For example, in some extreme cases where the absolute values of all elements in the pivot row are either extremely small or greater than 2 (such as $10^{-10}$ and 2.01, this can be commonly seen in a scaled spatial correlation matrix), the extremely small value is closer to 1 than the number just greater than 2, and therefore will be chosen as the pivot, possibly causing overflow. If using the maximal absolute value as pivot, however, overflow can be avoided.

On the other hand, when there is an extremely large number on the pivot row, it may seem like choosing a large number might potentially cause underflow. Here we provide the reason and a simple example which shows that such scenario will not affect the matrix condensation at all. When the pivot is very large, the entire $l^{th}$ column that is divided by the pivot will be close to 0. We can observe from Equation (2) and Equation (3) that when both $a_{i,l}$ and $a_{i+1,l}$ are 0 and $a_{k,l}$ is 1, the resulting element of $B^*$ will remain unchanged from $A$, which will not cause any problem.

For example, suppose $A$ is a 3x3 matrix as shown in Figure 3, and pivot $e$ is very large such that the remaining elements of the column become 0 after factoring out $e$. After applying Equation (2) to condense $A^*$, the final elements of $B^*$ are exactly the same as elements of $A$ with the pivot row and column taken out.

$$\underset{A}{Det\left(\begin{bmatrix} a & b & c \\ d & e & f \\ h & i & j \end{bmatrix}\right)} = e \cdot \underset{A^*}{Det\left(\begin{bmatrix} a & 0 & c \\ d & 1 & f \\ h & 0 & j \end{bmatrix}\right)} = e \cdot Det\underset{B^*}{\begin{bmatrix} Det\left(\begin{bmatrix} a & 0 \\ d & 1 \end{bmatrix}\right) Det\left(\begin{bmatrix} 1 & f \\ 0 & c \end{bmatrix}\right) \\ Det\left(\begin{bmatrix} 1 & d \\ 0 & h \end{bmatrix}\right) Det\left(\begin{bmatrix} 1 & f \\ 0 & j \end{bmatrix}\right) \end{bmatrix}} = e \cdot \underset{B^*}{Det\left(\begin{bmatrix} a & c \\ h & j \end{bmatrix}\right)}$$

Figure 3 Illustration of Large Pivot Value not Affecting Condensation Process

## 2.3 Row-wise or column-wise pivot factoring

After choosing the pivot, we need to decide whether to factor the pivot by row or by column. It turns out that for matrix condensation, both methods are the same mathematically. We can see from equation (2) that the $B_{ij}$ will always require the product of the pivot row counterpart and pivot column counterpart (Figure 4). Due to the associative property of multiplication it does not matter whether the multiplier $(1/a_{k,l})$ goes on the pivot row or pivot column.

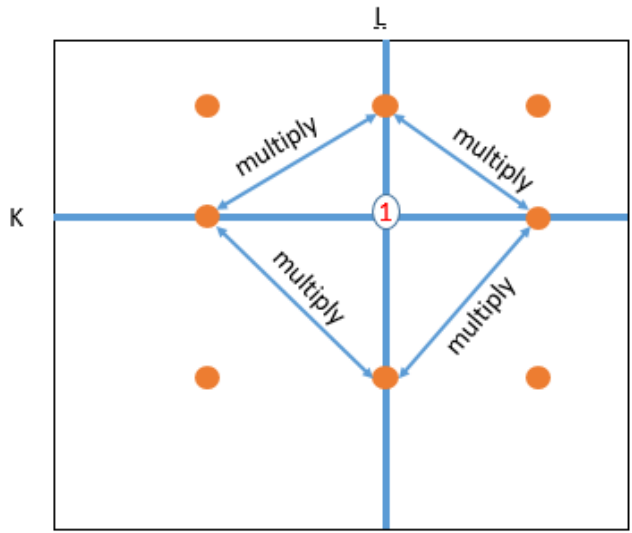

Figure 4 Visualization of Multiplication Pattern of Equation (2)

However, for a parallel implementation, in order to minimize communication it is better to factor the pivot within a single processor. For example, for storage using row-major order where the blocks distributed to the processors contain entire rows, we should factor the rows.

## *2.4 Improve CPU performance*

From a practical perspective, it is better to work on memory in-place rather than copying memory during the condensation process. When working in-place, however, during each step of partial pivoting one column of matrix $A$ is no longer used in the representation of matrix $B$, and thus the elements of $B$ are no longer truly contiguous in memory, which will introduce unnecessary CPU cache misses as the process proceeds with large matrices.

In order to overcome this, once we select the pivot row and column, we will switch the last column with the pivot column in memory as is shown in Figure 5. Column exchange will not affect the absolute value of the determinant, but if one is interested in the sign then the number of exchanges needs to be recorded. Column exchange makes the elements of $B$ in memory continuous again.

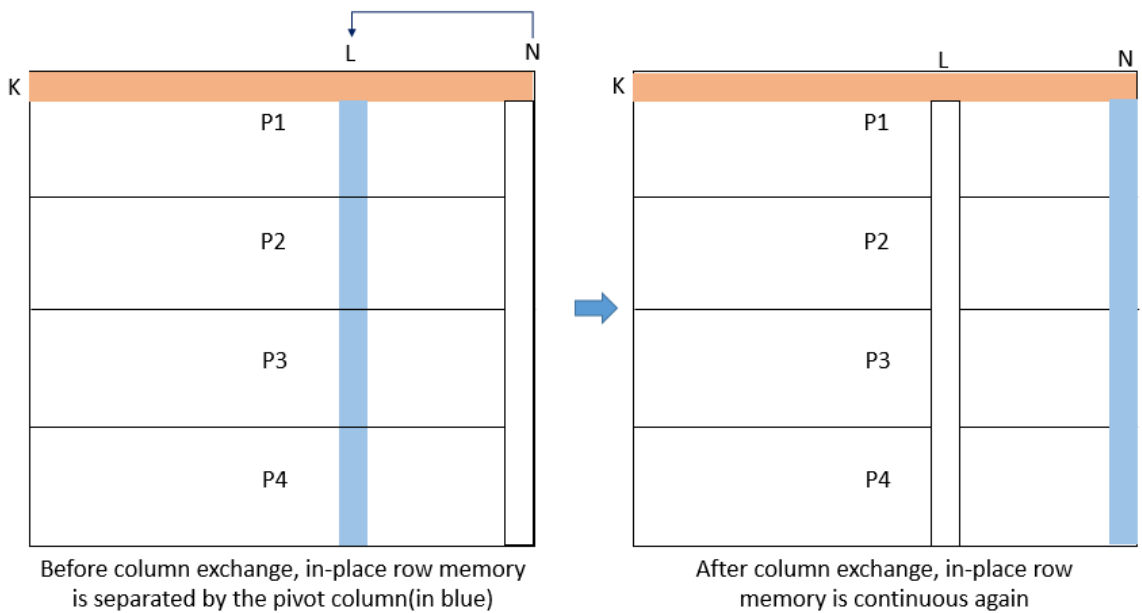

*Figure 5 Column exchange to make memory continuous*

## 2.5 Matrix Condensation V.S. Gaussian Elimination

Before doing a comparison, it is worth it to mention that most parallel linear algebra libraries (such as ScaLAPACK) use Block Gaussian Elimination for LU decomposition and determinant calculation. These block operations divide the large matrix into a series of submatrices and replace real number arithmetic with matrix operation for the purpose of reducing data communications. For a fair comparison, we will not consider block operations for both algorithms and we will leave block matrix condensation for future study. Also note that the pivoting referred to is partial pivoting and that the comparison is made on a distributed memory architecture.

As stated in [1], if Matrix Condensation starts from top to bottom, then the amount of floating point operations is exactly the same as Gaussian Elimination without partial pivoting. However, there are two unique properties of Matrix Condensation that, once utilized, can provide great advantages over Gaussian Elimination.

1. As mentioned in Section 2.1, Matrix Condensation can start with an arbitrary row whereas Gaussian Elimination has to start from one side to the other. Arbitrary pivoting row allows Matrix Condensation to utilize a more efficient data distribution scheme.

2. Once the pivot row is selected, Matrix condensation can do the pivoting within a single processor. On the contrary, Gaussian Elimination has to find a pivot across all rows and exchange data about the pivot between processors, which incurs additional communication time. Therefore, Matrix Condensation's unique pivoting property can save a significant amount of operations as pivoting can be achieved locally.

*Table 1 Gaussian Elimination V.S. Matrix Condensation*

| Metrics | Matrix Condensation | Gaussian Elimination |
|---|---|---|
| Computational Complexity | $O(N^3)$ | $O(N^3)$ |
| No Pivoting Communication | ✓ | ✗ |
| High Data Distribution Efficiency | ✓ | ✗ |
| Load Balanced | ✓ | ✓ |

The comparison between Matrix Condensation and Gaussian Elimination is summarized in Table 1 above. Matrix Condensation demonstrates advantages as it has higher data distribution efficiency and more convenient pivoting.

## 2.6 Pseudocode

The following pseudocode is an MPI-like implementation that assumes row-major storage. *N* is the matrix size and *n* is the number of processors. *A* is the large matrix and *myrank* is the variable containing current processor id. The pseudocode assumes *N* is divisible by *n*, also note that when the sign of determinant does not matter, step *4.11* is a very compact form of equation (2).

```
Input: N, A∈ R^{N×N}, n
Output: Logdet
1. Local_Logdet = 0, Logdet = 0
2. Local_Nrow = N/n, N_col = N
3. Scatter A among n processors and store them in Local_A
4. For i from 1 to Local_Nrow-1
      For p from 1 to n
        If myrank == p
          4.1 pivot_row = local_A[i,1:N_col]
          4.1 pivot = max(abs(pivot_row)), get pivot column index as j
          4.3 pivot_row = pivot_row/pivot
          4.4 pivot_row[j] = pivot_row[Ncol]
          4.4 local_Logdet = Local_Logdet+log(abs(pivot))
          4.5 row_shift = 1
        Else
          4.6 row_shift = 0
        End
        4.7 Broadcast pivot_row and j from processor p to all other processors
        4.8 pivot_column = local_A[i+row_shift:LocalNrow, j]
        4.9 local_A[i+row_shift: Local_Nrow, j] = local_A[i+row_shift: Local_Nrow, N_col]
        4.10 N_col = N_col -1
        For row = i+row_shift : N_row
          For col = 1: N_col
            4.11 local_A[row,col] = local_A[row,col] – pivot_column[row]*pivot_row[col]
          End
        End
      End
    End
  End
5. Gather local_A[Local_Nrow,1:N_col] ∈ R^{1×n} into global_A∈ R^{n×n}
6. Sum up all Local_Logdet and store in Logdet at master node
7. Do Gaussian elimination to global_A on master node
8. Logdet = Logdet +Logdet(global_A) on master node
9. Return Logdet
```

(Steps 4 through End are marked as "Parallel")

*Figure 6 Parallel Matrix Condensation Pseudo Code*

## 3. Experiments and Discussion

We stated in Section 2 that Matrix Condensation has certain advantages over Gaussian Elimination. In this section, we implement both algorithms and test them experimentally. In addition, we will also compare their performances with ScaLAPACK, a widely used parallel linear algebra packages.

All algorithms are implemented in Fortran and parallelized by MPI. Each algorithm is tested on dense matrices with size ranging from 1000x1000 to 8000x8000. We also test each matrix size with 1,2,4,8,16,32,64 and 128 processors for 5 individual runs. The element within each matrix is a 64-bit double precision number. All programs are executed on OU Supercomputing Center for Education & Research (OSCER), owned by the University of Oklahoma. The hardware specifications are described in Table 2. Note that each computing node contains 20 processor and the communication speed within a node is faster than communication across nodes.

Table 2 Hardware specifications of our experiment

| | Hardware specifications for the experiment |
|---|---|
| Server | Dell PowerEdge 13th-generation rack-mounted |
| CPU | dual Intel Xeon "Haswell" E5-2650 v3 2.3 GHz, with 9.6 Gigatransfers per second connection to RAM |
| Nodes | 20 CPU cores, 32 GB RAM, theoretical peak speed per node: 640 GFLOPs |
| RAM speed | 2133 MHz |
| Disk drive | single 1 TB SATA 7200 RPM |
| Networks | Infiniband for MPI call, Ethernet for rest |
| Storage | NFS storage over Ethernet |

For the two algorithms that we implemented ourselves (Matrix Condensation and Gaussian Elimination), we measure the total execution time, data distribution time and communication time on the master processor. The total execution time is measured from the end of data distribution till the end of all calculations. The data distribution time is the amount of time used to distribute the matrix $A$ among all nodes. Communication time measures the amount of time taken by MPI function calls, which includes the broadcasting, global pivoting and variable reduction operations. For the Gaussian Elimination using ScaLAPACK, only total execution time is measured and the block size is set to 1 to ensure fair comparison.

The final result is generated by averaging all 5 independent runs. We also make sure that all programs can accurately report the correct log determinant to at least 10 significant digits. The average execution time for all matrix sizes and different number of processors are shown in Table 3. All raw outputs and analysis scripts are available at the source code link.

Table 3 Average experiment run time for all algorithms

| Matrix Size | Matrix Condensation Execution time (s) After Averaging 5 runs | | | | | | | |
|---|---|---|---|---|---|---|---|---|
| | Number of Processors | | | | | | | |
| | 1 | 2 | 4 | 8 | 16 | 32 | 64 | 128 |
| 1000x1000 | 0.20 | 0.10 | 0.06 | 0.04 | 0.03 | 0.55 | 1.43 | 3.08 |
| 2000x2000 | 1.93 | 0.98 | 0.43 | 0.24 | 0.15 | 0.64 | 1.49 | 3.15 |
| 3000x3000 | 8.41 | 4.52 | 1.84 | 1.12 | 0.82 | 0.85 | 1.68 | 3.29 |
| 4000x4000 | 20.95 | 11.38 | 5.27 | 3.62 | 3.06 | 1.46 | 1.96 | 3.54 |
| 5000x5000 | 41.53 | 22.64 | 10.95 | 7.72 | 6.76 | 3.25 | 2.36 | 3.94 |
| 6000x6000 | 71.92 | 39.43 | 19.42 | 13.78 | 12.39 | 6.16 | 3.73 | 5.21 |
| 7000x7000 | 114.48 | 62.67 | 31.15 | 22.41 | 20.07 | 10.15 | 5.58 | 5.98 |
| 8000x8000 | 170.80 | 94.06 | 46.85 | 33.73 | 31.28 | 15.74 | 8.16 | 7.17 |

| Matrix Size | Gaussian Elimination Execution time (s) After Averaging 5 runs | | | | | | | |
|---|---|---|---|---|---|---|---|---|
| | Number of Processors | | | | | | | |
| | 1 | 2 | 4 | 8 | 16 | 32 | 64 | 128 |
| 1000x1000 | 0.19 | 0.11 | 0.07 | 0.05 | 0.04 | 3.62 | 9.29 | 11.14 |
| 2000x2000 | 1.83 | 0.98 | 0.44 | 0.26 | 0.18 | 4.15 | 15.54 | 22.99 |
| 3000x3000 | 8.10 | 4.47 | 1.85 | 1.16 | 0.89 | 4.56 | 19.07 | 35.10 |
| 4000x4000 | 20.37 | 11.28 | 5.27 | 3.66 | 3.15 | 5.30 | 21.97 | 44.36 |
| 5000x5000 | 40.51 | 22.39 | 10.92 | 7.79 | 6.99 | 7.25 | 24.27 | 55.07 |
| 6000x6000 | 70.36 | 39.17 | 19.41 | 13.94 | 12.53 | 10.23 | 26.38 | 59.73 |
| 7000x7000 | 111.93 | 62.31 | 31.03 | 22.70 | 21.19 | 14.40 | 29.24 | 67.77 |
| 8000x8000 | 167.44 | 94.22 | 47.08 | 34.27 | 31.76 | 19.49 | 31.90 | 75.15 |

| Matrix Size | Gaussian Elimination Using ScaLAPACK (Blocksize = 1) Execution time (s) After Averaging 5 runs | | | | | | | |
|---|---|---|---|---|---|---|---|---|
| | Number of Processors | | | | | | | |
| | 1 | 2 | 4 | 8 | 16 | 32 | 64 | 128 |
| 1000x1000 | 1.62 | 1.19 | 1.01 | 0.94 | 1.00 | 1.33 | 1.69 | 2.35 |
| 2000x2000 | 7.62 | 5.39 | 4.15 | 3.54 | 3.36 | 3.94 | 4.45 | 5.30 |
| 3000x3000 | 21.05 | 14.13 | 10.28 | 8.47 | 7.78 | 8.52 | 8.94 | 10.16 |
| 4000x4000 | 43.22 | 28.52 | 19.91 | 16.66 | 15.18 | 15.20 | 15.42 | 17.04 |
| 5000x5000 | 77.98 | 49.32 | 33.99 | 28.24 | 25.79 | 24.48 | 24.33 | 25.58 |
| 6000x6000 | 126.19 | 77.70 | 52.23 | 43.02 | 39.37 | 37.09 | 34.86 | 36.54 |
| 7000x7000 | 187.75 | 115.55 | 76.59 | 62.27 | 56.83 | 50.59 | 49.91 | 49.05 |
| 8000x8000 | 268.68 | 163.06 | 105.80 | 85.33 | 78.47 | 69.44 | 64.60 | 63.64 |

Given a problem size, speed-up is calculated by $T_s/T_p$, where $T_s$ is the fastest serial time among all three algorithms and $T_p$ is the execution time using p processors on the same problem size for specific algorithm. The speed-up of three algorithms for all matrix sizes is shown in Figure 7, the average speedup is calculated by averaging the speed-up across all problem sizes and it is shown in Figure 8.

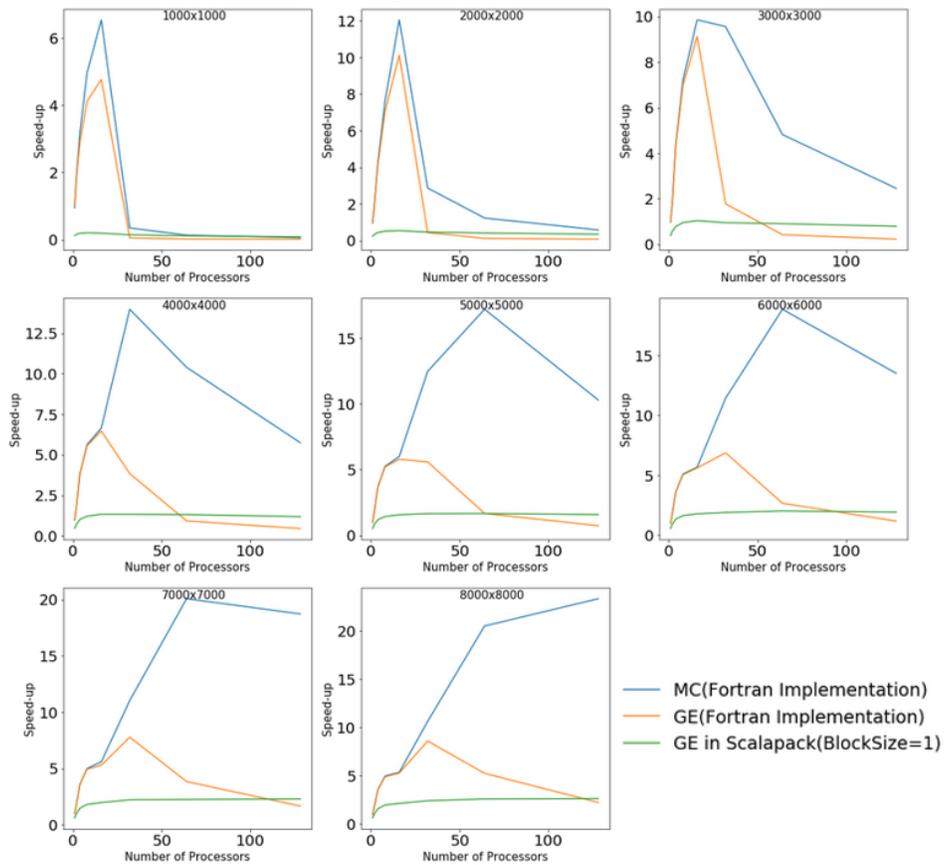

*Figure 7 Speed-up of all algorithms for different matrix sizes*

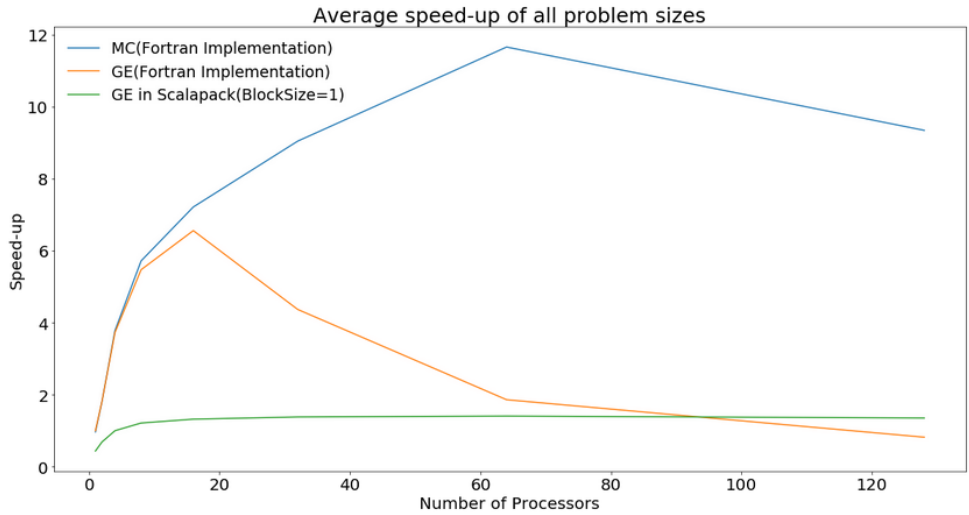

*Figure 8 Average speed-up for three algorithms*

We can see from Figure 7 and Figure 8 that ScaLAPACK with block size 1 has the lowest speed-up among all algorithms as its full power not being fully utilized due to the block size restriction.

For the other two self-implemented algorithms, Matrix Condensation demonstrates a better speed-up than Gaussian Elimination on all matrix sizes because of its better data distribution scheme and lower communication cost. In order to further support the above statement, we calculated the average data distribution time and communication time across all problem sizes for the two algorithms.

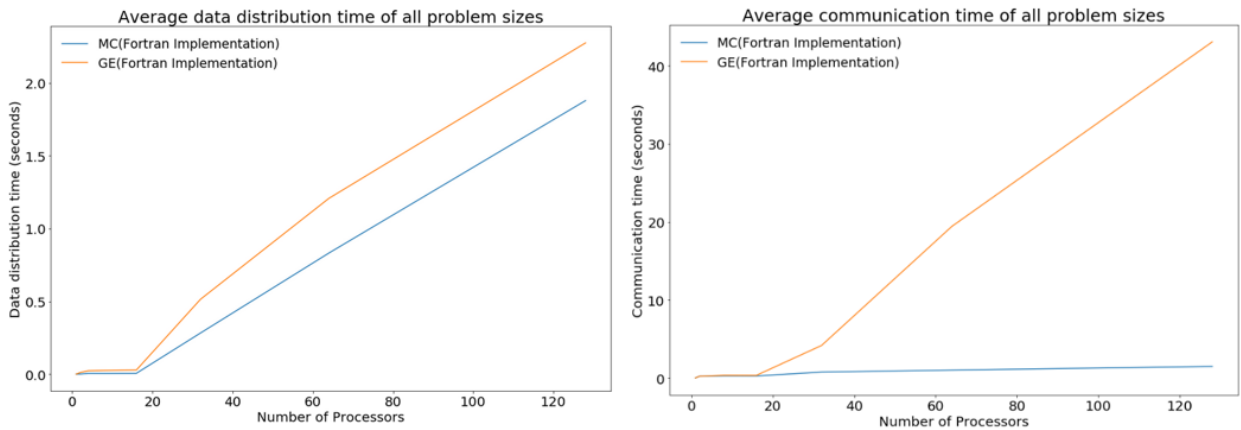

*Figure 9 Average data distribution time and communication time for MC and GE*

As shown in left of Figure 9, once the number of processors are above 20, the overall data distribution time increases almost linearly with number of processors for both algorithms. Nevertheless, Matrix Condensation always uses less distribution time than Gaussian Elimination, which indicates that block distribution is internally more efficient than cyclic distribution. Right side of Figure 9 shows that Matrix Condensation spends much less time than Gaussian Elimination on MPI function calls. This is due to the fact that Matrix Condensation is able to do the pivoting within processor whereas Gaussian Elimination has to do pivoting across processors. The slight variation between the two algorithms made a significant difference that ultimately gives Matrix Condensation great advantages in the execution time, which confirms the points in Section 2.5.

## Future Work

There can be many improvements on the current matrix condensation algorithm. First, most parallel linear algebra libraries uses block operations to save communication. It would be interesting to see how block operations can affect Matrix Condensation operations and whether Matrix Condensation can still demonstrate the advantages shown in this paper. Next, our methodology mainly deals with dense matrices, though the current algorithm could also be used to calculate the determinant of sparse matrices. The performance when handling sparse matrices could be further increased if one could capitalize on compact storage patterns of sparse matrices such that redundant operations are efficiently avoided. Lastly, it would be also worthwhile to extend our workflow to GPU devices and see the performance gain there.

## Conclusion

When it comes to determinant calculation, Gaussian Elimination has drawn most people's attention whereas the significance of matrix condensation has been more or less omitted over the years. In this paper we demonstrated that Matrix Condensation has many advantages over Gaussian Elimination in parallel computing, such as better data distribution and a better pivoting scheme. Matrix Condensation has great potential to surpass Gaussian Elimination in determinant calculation. Future studies are encouraged to fully demonstrate the significance of Matrix Condensation on block operations, sparse matrix and GPU devices.